# Teaching Autonomous Driving Using a Modular and Integrated Approach

Jie Tang[1], Shaoshan Liu[2], Songwen Pei[3], Stéphane Zuckerman[4], Chen Liu[5], Weisong Shi[6], and Jean-Luc Gaudiot[7]

[1]South China University of Technology, China
[2]PerceptIn Inc, U.S.A.
[3]University of Shanghai for Science and Technology, China
[4]Université de Cergy-Pontoise, France
[5]Clarkson University, U.S.A.
[6]Wayne State University, U.S.A.
[7]University of California, Irvine, U.S.A.

## Abstract:

*Autonomous driving is not one single technology but rather a complex system integrating many technologies, which means that teaching autonomous driving is a challenging task. Indeed, most existing autonomous driving classes focus on one of the technologies involved. This not only fails to provide a comprehensive coverage, but also sets a high entry barrier for students with different technology backgrounds. In this paper, we present a modular, integrated approach to teaching autonomous driving. Specifically, we organize the technologies used in autonomous driving into modules. This is described in the textbook we have developed as well as a series of multimedia online lectures designed to provide technical overview for each module. Then, once the students have understood these modules, the experimental platforms for integration we have developed allow the students to fully understand how the modules interact with each other. To verify this teaching approach, we present three case studies: an introductory class on autonomous driving for students with only a basic technology background; a new session in an existing embedded systems class to demonstrate how embedded system technologies can be applied to autonomous driving; and an industry professional training session to quickly bring up experienced engineers to work in autonomous driving. The results show that students can maintain a high interest level and make great progress by starting with familiar concepts before moving onto other modules.*

## 1. Introduction

In recent years, autonomous driving has become quite a popular topic in the research community as well as in industry. However, the biggest barrier to the rapid development of this field is a very limited talent supply. This is due to several problems: first, autonomous driving is the complex integration of many technologies, making it extremely challenging to teach; second, most existing autonomous driving classes focus on one technology of the complex autonomous driving technology stack, thus failing to provide a comprehensive introduction; third, without good integration experiments, it is very difficult for the students to understand the interaction between different technology pieces.

To address these problems, we have developed a modular and integrated approach to teach autonomous driving. First, we developed a textbook, along with a series of multimedia video lectures, to describe the technologies used in autonomous driving into modules. For students interested in autonomous driving, this book provides a comprehensive overview of the whole autonomous vehicle technology stack. For practitioners, this book presents many practical techniques and a sufficient number of references for them to perform an effective, deeper exploration of one particular module. In addition, to help the students understand the interactions between different modules, we developed platforms for hands-on integration experiments. To the best of our knowledge, this is the first work exploring techniques of teaching autonomous driving.

Our teaching methodology starts with an overview of autonomous driving technologies, followed by different technology modules, and ends with integration experiments. Note that the order of the modules can be flexibly adjusted based on the students' backgrounds and interest levels. We have successfully applied this methodology to three different scenarios: an introduction to autonomous driving class for undergraduate students with limited technology background; a graduate level embedded systems class, in which we added a session on autonomous driving; as well as a two-week professional training for seasoned engineers.

The rest of the paper is organized as follows: section 2 reviews existing classes on autonomous driving; section 3 presents an overview of autonomous driving technologies; section 4 delves into the details of the proposed modular and integrated teaching methodology; section 5 presents the three case studies where we applied the proposed methodology; and we draw the conclusions in section 6.

## 2. Existing Classes on Autonomous Driving

Autonomous driving has been attracting the attentions from the academia as well as the industry. However, the comprehensive and complex autonomous driving system involves a very diverse set of technologies including sensing, perception, localization, decision making, real-time operating system, heterogeneous computing, graphic/video processing, and cloud computing, *etc.* It sets very high requirements for lecturers to deeply master all aspects of relative technologies. On the other side, it is even more a challenge for students to understand the interactions between these technologies.

Problem-based learning (PBL) is a feasible and practical way to teach relative knowledge and technologies of autonomous driving [20]. Costa built a simulator by integrating Gazebo 3D simulator for students to design and understand autonomous driving systems [13]. In terms of the educational methodology on Learning from Demonstration, Arnaldi proposed an affordable setup for machine learning applications of autonomous driving by implementing embedded programming for small autonomous driving cars [14]. However, these approaches usually cover only one or two technologies, such as machine learning, and fail to provide a comprehensive understanding of the whole system.

Several major universities already offer autonomous driving related classes. For instance, MIT offers two courses on autonomous driving. The first one focuses on Artificial Intelligence and is available online to the public and registered students. This course invites several guest speakers on the topic of deep learning, reinforcement learning, robotics, psychology, *etc* [15]. The other concentrates on Deep Learning for Self-Driving Cars and teaches deep learning knowledge by building a self-driving car [16]. Stanford also has a course for introducing the key artificial intelligence technologies that could be used for autonomous driving [17]. However, these classes all focus on machine learning and do not provide a good coverage on different technologies involved in autonomous driving. It is thus hard for students to get a comprehensive understanding of autonomous driving systems.

On experimental platforms and competition developments, Paull *et al.* proposed Duckietown which is an open-source and inexpensive platform for autonomy education and research [18]. The autonomous vehicles are equipped with Raspberry Pi 2 and a monocular camera for sensing. In addition, there are several competitions of autonomous driving for broad levels students to stimulate the passion and inspiration of learning and spurring key technologies in autonomous cars and related traffic systems [19, 21-23]. However, to use these platforms and to enter these competitions, students need to first gain a basic understanding of the technologies involved as well as their interactions; this type of education is currently lacking.

As seasoned autonomous driving researchers and practitioners, we think the best way to learn how to create autonomous vehicle systems is to first grasp the basic concepts in each technology module, and then to integrate these modules to understand how these modules interact.  Existing autonomous driving classes either focus on only one or two technologies, or directly have the students to build a working autonomous vehicle. As a result, this disconnection between individual technology module and system integration creates a very high entry barrier for students interested in autonomous driving, and often scare interested students away from entering this exciting field. To address this exact problem, in this paper we present our modular and integrated approach and share our experiences with autonomous driving education.

# 3. Overview of Autonomous Driving Technologies

Autonomous driving is not one single technology but rather a complex system integrating many technologies [1]. In this section, we provide a brief overview of the technologies used in autonomous driving.  As we will demonstrate, it is technically challenging to teach the details of each technology involved, not to mention the integration of all technologies.  Therefore, it is extremely difficult to provide a comprehensive education on autonomous driving unless the correct methodology is in place.

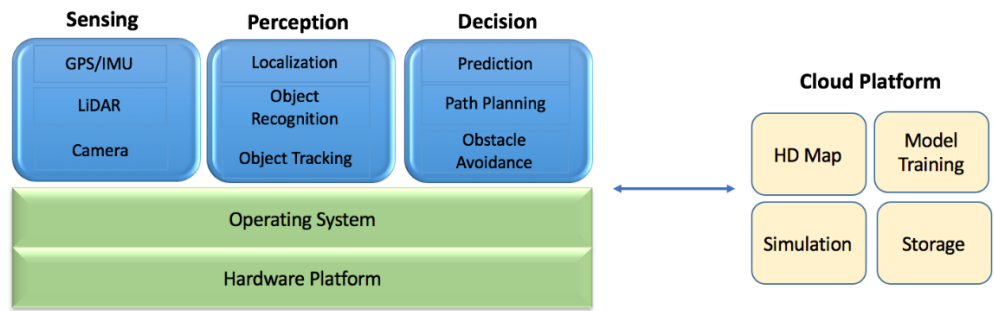

Figure 1: autonomous driving technologies

The autonomous driving technology stack consists of three major subsystems: algorithms, including sensing, perception, and decision; client, including the operating system and hardware platform; and the cloud platform, including data storage, simulation, high-definition (HD) mapping, and deep learning model training.  The algorithm subsystem extracts meaningful information from sensor raw data to understand its environment and make decisions about its actions. The client subsystem integrates these algorithms to meet real-time and reliability requirements. For instance, if the sensor camera generates data at 60 Hz, the client subsystem needs to make sure that the longest stage of the processing pipeline takes less than 16 milliseconds (ms) to complete. The cloud platform provides offline computing and storage capabilities for autonomous cars. Using the cloud platform, we are able to test new algorithms and update the HD map—plus, train better recognition, tracking, and decision models.

## 3.1 Sensing

A variety of sensors are used in autonomous driving.  Each is an independent research field:

- *GNSS receivers*, especially those with real-time kinematic (RTK) capabilities, help autonomous vehicles localize themselves by updating global positions with at least meter-level accuracy.
- *LiDAR* is usually used for creating High-Definition (HD) maps, real-time localization, as well as obstacle avoidance. LIDAR works by bouncing a laser beam off of surfaces and measuring the reflection time to determine distance. A typical LiDAR unit used in autonomous vehicles covers a range of 150 meters and takes over 1 million spatial points (with <x, y, z> coordinates) per second. Each point is associated with a reflectivity attribute, which can be used to identify points between different frames. By comparing the displacements of the spatial points between two frames, we can derive the displacement of the vehicle.
- *Cameras* are mostly used for object recognition and tracking tasks, such as lane detection, traffic light detection, pedestrian detection. Existing implementations usually mount multiple cameras around the vehicle to detect, recognize, and track objects. These cameras usually run at 60 Hz, and, when combined, can generate over 1 GB of raw data per second.
- The *radar and sonar* sub-system is used as the last line of defense in obstacle avoidance. The data generated by radar and sonar shows the distance from the nearest object in front of the vehicle's path. When an object is detected as not far ahead and there is danger of a collision, the autonomous vehicle should take evasive actions.

## 3.2 Perception

Perception is about *understanding the environment*, including the current location, object recognition, and tracking *etc*. Several different techniques can be utilized to acquire accurate vehicle position updates in real-time. The natural choice for localization is to use *GNSS* directly. Nonetheless, we cannot solely rely on this because of the multipath problems, meaning that signals may bounce off of buildings, introducing noise. We must thus have redundant perception mechanisms:
- *LiDAR* is commonly used in localization. First the localization sub-system extracts point clouds through LiDAR scans, providing a geometry description of the environment. Then, the localization sub-system compares a specific observed shape against the shapes in a confined region of the HD map to reduce uncertainty and tracks the positions of the moving vehicle.
- *Cameras* can be used for localization. The technology used is called visual odometry. Visual odometry works by first extracting spatial points through stereo vision and then by comparing the locations of the detected spatial points between consecutive frames to deduce the movement of the vehicle between the two frames.
- Object recognition and tracking can be achieved with *Deep Learning which,* in recent years has seen a rapid development. It implements accurate object detection and tracking using camera inputs. A convolution neural network (CNN) is a type of deep neural network that is widely used in object recognition tasks. A general CNN evaluation pipeline usually consists of the following layers: 1) the convolution layer which uses different filters to extract different features from the input image; 2) the activation layer which decides whether to activate the target neuron or not; 3) the pooling layer which reduces the spatial size of the representation to reduce the number of parameters; and last, 4) the fully connected layer which connects all neurons to all activations in the previous layer. Once an object is identified, object tracking technology can be used to track nearby moving vehicles, as well as pedestrians crossing the road to ensure the current vehicle does not collide with moving entities.

## 3.3 Decision Making

In the decision making stage, action prediction and path planning mechanisms are combined to *generate an effective action plan* in real time. The main challenge of autonomous driving planning is to make sure that autonomous vehicles travel safely in complex traffic environments. The decision making unit generates predictions of nearby vehicles before deciding on an action plan based on these predictions. To predict the actions of other vehicles, one can generate a stochastic model of the reachable position sets of the other traffic participants and associate these reachable sets with probability distributions.

Planning the path of an autonomous vehicle in a dynamic environment is a complex problem, especially when the vehicle is required to use its full maneuvering capabilities. One approach is to search all possible paths and utilize a cost function to identify the best path. However, this requires enormous computational resources which may make it incapable of delivering real-time navigation plans. To circumvent this computational complexity and provide effective real-time path planning, probabilistic planners can be utilized.

## 3.4 Client System

The client system integrates the above-mentioned algorithms together to meet real-time and reliability requirements. There are three challenges to overcome: 1) the system needs to ensure that the processing pipeline is sufficiently fast to consume – and process - the enormous amount of sensor data generated; 2) if a part of the system fails, it needs to be robust enough to recover from the failure; and 3), the system needs to perform all the computations under energy and resource constraints [2].

Specifically, the above mentioned components exhibit very different behaviors. For example, the planning and control algorithms stress the CPU, the object recognition and tracking algorithms stress the GPU, while the HD maps stress the memory. Therefore, the challenge is to design a computing hardware system which addresses these demands, all within limited computing resources and power budget.

| Application Layer | Sensing | Perception | Decision | Other |
|---|---|---|---|---|
| Operating System Layer | ROS Node | ROS Node | ROS Node | ROS Node |
| Run-Time Layer | Execution Run-Time ||||
| | OpenCL ||||
| Computing Platform | I/O Sub system | CPU |||
| | | Shared Memory |||
| | | DSP | GPU | FPGA |

Figure 2: autonomous driving client system

Figure 2 shows the complexity of the client system. At the level of the computing platform layer, a heterogeneous architecture consists of an I/O subsystem that interacts with the front-end sensors; a DSP to pre-process the image stream to extract features; a GPU to perform object recognition and some other deep learning tasks; a multi-core CPU for planning, control, and interaction tasks; an FPGA that can be dynamically reconfigured and time-shared for data compression and uploading, object tracking, and traffic prediction, *etc*. These computing and I/O components communicate through a shared memory. On top of the computing platform layer, a run-time layer maps different workloads to the heterogeneous computing units and schedule different tasks at. On top of the Run-Time Layer, an Operating Systems layer encapsulates each task in autonomous driving and facilitates their communications.

## 3.5 Cloud Platform for Autonomous Driving

Autonomous driving clouds provide essential services to support autonomous vehicles. These services include but not limited to distributed simulation tests for new algorithm deployment, offline deep learning model training, and High-Definition (HD) map generation. These services require infrastructure support including distributed computing, distributed storage, as well as heterogeneous computing [3].

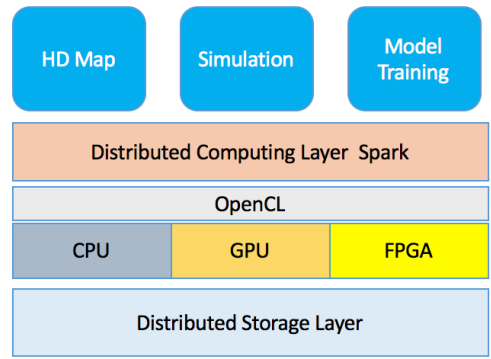

Figure 3: autonomous driving cloud platform

More specifically, HD maps have many layers of information. At the bottom layer, instead of using satellite imagery, a grid map is generated by raw LiDAR data, with a grid granularity of about 5 centimeters by 5 centimeters. This grid basically records elevation and reflection information of the environment in each cell. On top of the grid layer, there are several layers of semantic information. For instance, lane information is added to the grid map to allow autonomous vehicles determine whether they are on the correct lane when moving. On top of the lane information, traffic sign labels are added to notify the autonomous vehicles of the local speed limit, whether traffic lights are nearby, *etc*. This gives an additional layer of protection in case the sensors on the autonomous vehicles fail to catch the signs. Traditional digital maps have a refresh cycle of six to twelve months. However, to make sure the HD maps contain the most up-to-date information, the refresh cycle for HD maps is shortened to one week. Thus, HD map creation and maintenance puts tremendous stresses on the cloud platform.

# 4. Modular and Integrated Teaching Approach

As demonstrated in the previous section, autonomous driving involves many technologies, and thus making autonomous driving education extremely challenging. To address this problem, we propose a modular and integrated teaching methodology for autonomous driving. To implement this methodology, we have developed modular teaching materials including a textbook and a series of multimedia online lectures, as well as integration experimental platforms to enable a comprehensive autonomous driving education.

## 4.1 Teaching Methodology

In the past two years, we have taught undergraduate and graduate level university classes as well as brought up new engineers on autonomous driving. A common problem that we found was that the first encounter on this subject usually scared away a lot of students since the students felt that autonomous driving was too complicated for them. Similarly, when bringing up experienced engineers to get into the field of autonomous driving, these engineers felt it was extremely stressful to get out of their comfort zone, especially as the subject touches upon many new areas.

On the other hand, we found that a modular and integrated approach is an effective way of teaching autonomous driving. Using this approach, we first break the complex autonomous driving technology stack into modules and have the students to start with their familiar module, and then have them move on to learn other modules. This allows students to maintain a high interest level and make satisfactory progress throughout the learning process. Once the students have gone through all the modules, they are challenged to perform a few integration experiments to understand the interactions between these modules. Besides being an effective teaching method, this approach allows the instructors to flexibly adapt the class curriculum to the needs of students with different technology backgrounds, including undergraduate students with little technology background, graduate students with a general computer science technology background, and seasoned engineers who are experts in a particular field.

Figure 4 illustrates the proposed modular and integrated teaching approach: the class is divided into nine modules and integration experiments. Undergraduate and graduate students can both start with a general *overview* of autonomous driving technologies, where the undergraduate students may need more time to understand the basics of the technologies involved. Then, they can move on to *localization*, followed by *traditional perception* and *perception with deep learning*. Next, they can learn about the decision-making pipeline, including *planning and control*, *motion planning*, as well as *end-to-end planning*. Once the students are done with these, they can delve into *client systems* and *cloud platforms*. At last, they can perform integration experiments to understand the interactions between these modules.

On the other hand, seasoned engineers with embedded system background can start with the general *overview* and then directly move on to the *client systems* module such that they can get familiar with the new materials from the perspective of their comfort zone, and thus allowing them to maintain a high interest level. They can then move on to the *cloud platform*, which also focuses on system design and still within their comfort zone. Once they master these modules, they are equipped with enough background knowledge as well as confidence to learn the rest of the modules.

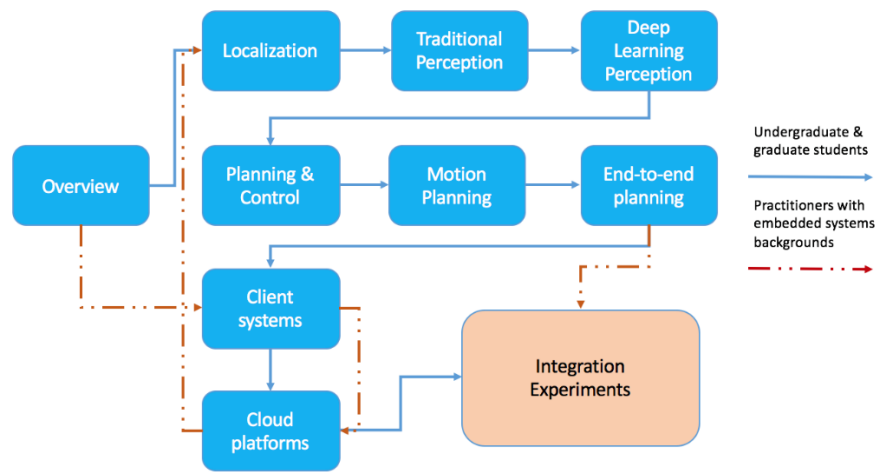

Figure 4: example of modular and integrated teaching approach

## 4.2 Modular Teaching Materials

First, as shown in Figure 5, to cover all the major modules in the autonomous driving technology stack, we have developed a textbook, *Creating Autonomous Vehicle Systems* [4]. This is the first technical overview of autonomous vehicles where we share our practical experiences of creating autonomous vehicle systems. This book consists of 9 chapters: chapter 1 provides an overview of autonomous vehicle systems; chapter 2 focuses on localization technologies; chapter 3 discusses traditional techniques used for perception; chapter 4 discusses deep learning based techniques for perception; chapter 5 introduces the planning and control sub-system, especially prediction and routing technologies; chapter 6 focuses on motion planning and feedback control of the planning and control subsystem; chapter 7 introduces reinforcement learning-based planning and control; chapter 8 delves into the details of client systems design; and chapter 9 provides the details of cloud platforms for autonomous driving.

This book is aimed at students, researchers and practitioners alike. For undergraduate or graduate students interested in autonomous driving, this book provides a comprehensive overview of the whole autonomous vehicle technology stack. For autonomous driving practitioners, this book presents many practical techniques in implementing autonomous driving systems. For researchers, this book provides plenty of references for an effective, deeper exploration of the various technologies.

Along with the textbook, in cooperation with IEEE Computer Society and O'Reilly, we have developed a series of online lectures to introduce each module [5, 6]. Along with the multimedia presentations, this allows students to easily acquire an in-depth understanding of a specific technology.

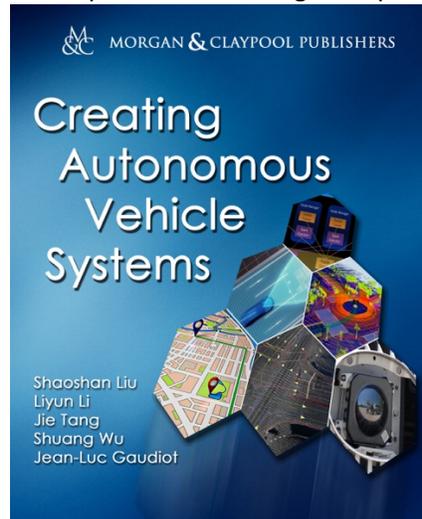

Figure 5: *Creating Autonomous Driving Vehicle Systems* Textbook

## 4.3 Integration Experimental Platforms

A common problem of autonomous driving education is the lack of an experimental platform. In most autonomous driving classes, people use simulators to verify the performance of newly developed algorithms [11]. Nonetheless, the simulation approach fails to provide an environment for students to understand the interaction between different modules. On the other hand, using an autonomous driving vehicle as an experimental platform is not practical due to the cost, as a demo autonomous vehicle can easily cost over $800,000.

A straightforward way to perform integration experiments is to use our mobile phones. Nowadays, mobile phones usually consist of many sensors (GPS, IMU, cameras, *etc.*) as well as powerful heterogeneous computing platforms (with CPU, GPU, and DSP). Therefore, mobile phones can be used as a great integration experimental platform for localization and perception tasks. For instance, as demonstrated in [8], we have successfully implemented real-time localization, obstacle detection and avoidance, as well as planning and control functions on a Samsung Galaxy 7 mobile phone to drive a mobile robot at 5 miles per hour.

Besides integrating autonomous driving algorithms on mobile phones, we have also developed an integration experimental platform to emulate autonomous vehicles. As shown in Figure 6, the platform consists of three layers: the bottom layer is the mobile robot chassis that can receive commands and send status through the serial interface. Like a car, the commands supported by the chassis include *accelerate, brake, turn left, turn right, maintain speed, etc*, and the status data returned include *wheel odometry* as well as current *orientation*. On top of the chassis is the computing layer where students can implement different modules in the main computing system. On top of the computing layer is the sensing layer. In this design, we use a stereo visual inertial module with computing capability [7]. This way, students can perform sensor fusion to merge wheel odometry, IMU, and computer vision data to derive the latest location of the vehicle.

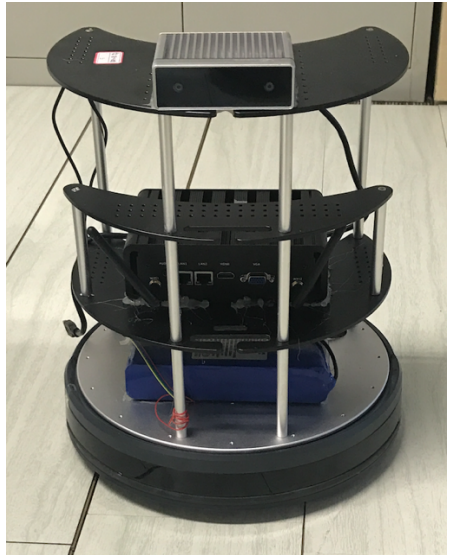

Figure 6: autonomous driving integration experimental platform

# 5. Case Studies

As shown in Table 1, several institutes and universities already have adopted the teaching methodology as well as the experimental platforms described in this paper. For instance, we are working closely with the Cyber Physical Systems program [25] and the Connected and Autonomous Driving (CAR) Laboratory at Wayne State University to develop an autonomous driving class in their program. Particularly, we have integrated the autonomous driving materials as well as the experimental platform shown in Figure 6 into the Wayne State University CSC5280, Introduction to Cyber-Physical Systems class, which will be offered in Fall 2018.

Table 1: teaching methodology adoption

| Institute/University | Class Name | Number of Students | Class Level |
|---|---|---|---|
| Clarkson University | Introduction to Autonomous Driving | 50 | undegraduate |
| University of Shanghai for Science and Technology | Advanced Autonomous Driving | 50 | undegraduate/graduate |
| South China University of Technology | Embedded Systems Design | 100 | undegraduate/graduate |
| IEEE Computer Society | Creating Autonomous Vehicle Systems | unlimited | professional online training |
| Wayne State University | Introduction to Cyber-Physical Systems | 50 | graduate |

In this section, we present three case studies of applying the aforementioned teaching methodology and materials: the first one is an introduction to autonomous driving class for undergraduate students with limited technology backgrounds; the second one is a graduate level embedded system class, in which we added a session on autonomous driving; and the last one is a two-week professional training for seasoned engineers. These three case studies were carefully selected to demonstrate the flexibility of using the proposed approach to teach autonomous driving.

## 5.1 Introduction to Autonomous Driving Class

We have developed an introduction to an autonomous driving class for undergraduate and graduate students. The class consists of 15 to 20 lectures depending on the length of the quarter or semester. Also, a 20-hour experiment session is required for integration experiments. The purpose of the class is to provide a technical overview on autonomous driving, for students with basic experience in programming, algorithms, and operating systems.

Due to their limited background, we do not expect the students to fully understand all modules, but we intend to maintain their interest level and equip them with the basic knowledge to delve into the modules that they are particularly interested in. To achieve this, we follow the approach shown in Figure 4 and divide the class into nine modules. To maintain a high level of interest, at the beginning of each session, we play a short video, such as [5], to provide a summary and demo of the technologies discussed. Then we move on to the details regarding how each technology is implemented. Also, throughout the class, we have the students use mobile phones to perform integration experiments. Specifically, for localization experiments, we first have the students extract real-time GPS localization data; then we have the students improve their localization data by fusing IMU data with GPS data. For perception, we have the students install a deep learning framework, such as MXNET [12], onto their mobile phones and run simple object detection networks.

Around 100 students took this class. Throughout the class, we closely monitored the students' interest level and performance, and we had two interesting observations: first, at the end of the class, we asked the students which was the module that they were most interested and would like to dig deeper. The results are summarized in Figure 7, indicating that students had diverse interests such that each module has similar interest level, with localization being the most popular module.

Second, we have developed multiple integration experiments for the students to gain deep understanding of the interaction between different modules. Since it is an introductory class, for the advanced integration experiments, such as fusing GPS, IMU and Camera data to provide accurate location updates in real time, we did not provide enough technical background in the lectures. Thus in order to accomplish these tasks, students not only needed to perform their own research to get enough technical background but also to spend significant time and effort to perform the experiments. Indeed, we did not expect any student to be able to accomplish the advanced tasks but to our surprise, 8% of the students were able to successfully accomplish these advanced tasks. These observations demonstrate that with a modular and integrated teaching approach, students not only get a great overview of the technologies but are also able to delve into the modules of their interests and become experts.

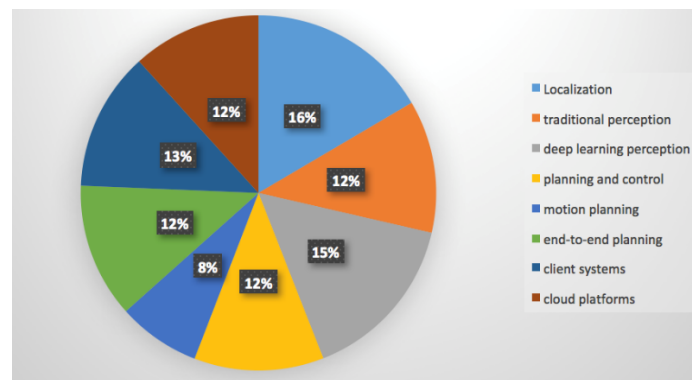

Figure 7: normalized interest level for each module

## 5.2 Adding Autonomous Driving Materials to Embedded Systems Class

We have also added a session in an existing graduate-level embedded systems class to explore how embedded systems technologies can be integrated in autonomous driving systems. The embedded systems class is a 20-week semester with 60 hours of lectures and 20 hours of experiments, in which we allocated 6 hours of lectures and 10 hours of experiments for autonomous driving contents. Interestingly, before we started this class, we checked with the students on whether they would start their engineering career in autonomous driving. Most students were highly interested but at the same time feared that autonomous driving was too complicated for them.

With the hope to ease the students' fears towards autonomous driving, we placed the autonomous driving session at the end of the class, after the students grasped the basic skills of designing and implementing embedded systems with different software and hardware optimization techniques, such as heterogeneous computing. Prior to starting this case study session, all students' understanding of autonomous driving was limited to the conceptual level. Out of the 56 students enrolled, only 10 students were able to list some technologies, such as localization and perception, involved in autonomous driving, but none understood the details of these technologies.

Due to the limited time available in the session, we first presented an overview on autonomous driving and then focused on two modules localization and perception. We delved into two very simple algorithm implementations, ORB-SLAM [9] for localization, and SqueezeNet [10] for object detection. Next we put students into groups of four to perform integration experiments with these algorithms on their Android cell phones and had them compare the performance of using CPU only *vs*. the performance of using heterogeneous computing components such as GPU and DSP. After completing the project, the students were asked to summarize their design choices and present their results in class. The presentations helped them understand the techniques used by other groups and they could learn from each other through the presentations.

The results were encouraging. First, it was very interesting to see different optimization strategies from different groups. Some groups prioritized computing resources for localization tasks to guarantee frequent position updates whereas other groups prioritized computing resources for perception to guarantee real-time obstacle avoidance. Second, through this session, the fears towards autonomous driving went away, the after class survey indicated that 85% of the students would like to continue learning autonomous driving.

## 5.3 Professional Training

For autonomous driving companies, one of the biggest challenges is the difficulty to recruit autonomous driving engineers since there is only a very limited talent pool with autonomous driving experiences. Therefore, it is key to develop a professional training session to quickly equip seasoned engineers with the technical knowledge to delve into one module of autonomous driving.

We worked closely with an autonomous driving company, to quickly bring up their engineers, mostly with embedded systems and general software engineering background. The challenges were three-fold: first, the training session was only two-week long, not enough time to delve into the technology details. Second, in two weeks' time we needed to place the engineers into different engineering roles although they came from similar technology backgrounds. Third, confidence was a big issue for these engineers as they were concerned as to whether they could handle the complexity of autonomous driving in a short amount of time.

To address these challenges, following the methodology presented in Figure 4, in the first week we had the engineers all start with the technology overview, followed by the client systems and the cloud platforms modules. Since the engineers came from embedded systems and general software engineering background, they were very comfortable to start with these modules. Through system modules, they learned about the characteristics of different workloads as well as how to integrate them on embedded and cloud systems. Then in the second week, based on the engineers' performance in the first week as well as their interest levels towards different technologies, we assigned them to dig deeper into a specific module, such as perception, localization, or decision making.

For integration experiments, unlike in the undergraduate or graduate classes, the engineers got the chance to work on the real product shown in Figure 8, after two weeks' training. In this training session, seven engineers were successfully on-boarded to the team, one was assigned to the sensing team, two were assigned to the perception team, two were added to the localization team, and two were assigned to the decision-making team. A demo video of the development process is shown in [24].

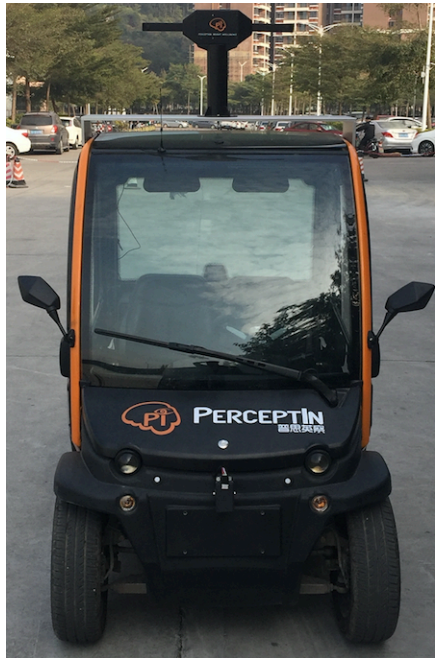
Figure 8: PerceptIn autonomous vehicle

# 6. Conclusions

We were often asked what the most important technology in autonomous driving was, our answer was always integration.  As mentioned above and we stress here again, autonomous driving is not one single technology but rather a complex system integrating many technologies. However, before integration happens, one has to understand each technology module involved. Existing autonomous driving classes often focus on one or two technology modules, or directly have the students build a working autonomous vehicle, thus creating high entry barriers for students. Surprisingly, most students are indeed highly interested in autonomous driving but it is the fear that they could not handle the complexities involved often drive them away.

   To address this problem, we have developed a modular and integrated approach to teach autonomous driving.  This approach breaks the complex autonomous driving system into different technology modules and have the students understand each module first.  After the students grasp the basic concepts in each technology module, we then have the students perform integration experiments to help them understand the interactions between these modules.

   We have successfully applied this methodology to three pilot case studies: an undergraduate-level introduction to autonomous driving class, a graduate-level embedded systems class with a session on autonomous driving, as well as a professional training session in an autonomous driving company. Although the students in these three pilot case studies have very diverse backgrounds, the modular teaching approach allowed us to adjust the order of modules flexibly to fit the needs of different students, and the integration experiments enabled the students to understand the interactions between different modules. This gave the students a comprehensive understanding of the modules as well as their interactions.  In addition, our experiences showed that the proposed approach allowed students start with their comfortable modules and then move on to other modules, therefore enabling the students to maintain a high interest level and good performance.

# References

1. Liu, S., Peng, J. and Gaudiot, J.L., 2017. Computer, Drive My Car! Computer, 50(1), pp.8-8.
2. Liu, S., Tang, J., Zhang, Z. and Gaudiot, J.L., 2017. Computer Architectures for Autonomous Driving. Computer, 50(8), pp.18-25.
3. Liu, S., Tang, J., Wang, C., Wang, Q. and Gaudiot, J.L., 2017. A Unified Cloud Platform for Autonomous Driving. Computer, (12), pp.42-49.
4. Liu, S., Li, L., Tang, J., Wu, S. and Gaudiot, J.L., 2017. Creating Autonomous Vehicle Systems. Synthesis Lectures on Computer Science, 6(1), pp.i-186.
5. IEEE Computer Society, Creating Autonomous Vehicle Systems, accessed 1 Feb 2018, https://www.youtube.com/watch?v=B8A6BiRkNUw&t=93s
6. OReilly, Enabling Computer-Vision-Based Autonomous Vehicles, accessed 1 Feb 2018, https://www.youtube.com/watch?v=89giovpaTUE&t=434s
7. PerceptIn, PerceptIn Ironsides Visual Inertial Computing Module, accessed 1 Feb 2018, https://www.perceptin.io/ironsides
8. PerceptIn, PerceptIn Robot System Running on a Cell Phone, accessed 1 Feb 2018, https://www.youtube.com/watch?v=Mib8SXacKEE
9. ORB-SLAM, accessed 1 Feb 2018, http://webdiis.unizar.es/~raulmur/orbslam/
10. SqueezeNet, accessed 1 Feb 2018, https://github.com/DeepScale/SqueezeNet
11. Tang, J., Liu, S., Wang, C., and Liu, C., 2017. Distributed Simulation Platform for Autonomous Driving. International Conference on Internet of Vehicles (IOV) 2017: pp. 190-200
12. Apache MXNET, accessed 1 Feb 2018, https://mxnet.apache.org/
13. Costa V., Rossetti R., and Sousa A., Simulator for Teaching Robotics, ROS and Autonomous Driving in a Competitive Mindset, International Journal of technology and human interaction, 13(4): 14, 2017.
14. Arnaldi N., Barone C., Fusco F., Leofante F., and Tacchella A., Autonomous Driving and Undergraduates: an Affordable Setup for Teaching Robotics, Proceedings of the 3rd Italian Workshop on Artificial Intelligence and Robotics, pp.5-9, Genova, Italy, November 28, 2016.
15. Artificial General Intelligence, accessed 1 Feb 2018, https://agi.mit.edu/
16. Deep Learning for Self-Driving Cars, accessed 1 Feb 2018, https://selfdrivingcars.mit.edu/
17. Artificial Intelligence: Principles and Techniques, accessed 1 Feb 2018, http://web.stanford.edu/class/cs221/
18. Paull L., Tani J., *et al*. Duckietown: an Open, Inexpensive and Flexible Platform for Autonomy Education and Research, IEEE International Conference on Robotics and Automation, Singapore, May. 2017, pp.1-8.
19. Karaman S., Anders A., Boulet M., et al. Project-based, collaborative, algorithmic robotics for high school students: Programming self-driving race cars at MIT, IEEE Integrated STEM Education Conference, pp.195-203, 2017.
20. Tan S., Shen Z., Hybrid Problem-Based Learning in Digital Image Processing: A Case Study, IEEE Transactions on Education, 2017, pp (99):1-9.
21. Robotica 2017, accessed 1 Feb 2018, http://robotica2017.isr.uc.pt/index.php/en/competitions/major/autonomous-driving
22. Autonomous Driving Challenge, accessed 1 Feb 2018, http://www.autodrivechallenge.org/
23. NXP CUP Intelligent Car Racing, accessed 1 Feb 2018, https://community.nxp.com/groups/tfc-emea

24. PerceptIn Autonomous Vehicle Development, accessed 1 Feb 2018, https://www.youtube.com/watch?v=rzRC57IXtRY
25. Wayne State University Cyber Physical Graduate Certificate Program, accessed 1 Feb 2018, http://engineering.wayne.edu/cyber/news/college-of-engineering-launches-new-graduate-certificate-program-in-cyber-physical-systems-28587